\documentclass[fleqn,12pt]{wlscirep}

\usepackage{color}
\usepackage{ulem}
\usepackage{soul}
\usepackage{multirow}
\usepackage{array}
\usepackage{amsmath}
\usepackage{amsfonts}
\usepackage{amssymb}
\usepackage{pdflscape}
\usepackage{afterpage}
\usepackage{caption}
\usepackage{xr}
\usepackage{float}
\DeclareMathAlphabet{\mathpzc}{OT1}{pzc}{m}{it}
\usepackage{lineno}
\newcolumntype{L}[1]{>{\raggedright\let\newline\\\arraybackslash\hspace{0pt}}m{#1}}
\newcolumntype{C}[1]{>{\centering\let\newline\\\arraybackslash\hspace{0pt}}m{#1}}
\newcolumntype{R}[1]{>{\raggedleft\let\newline\\\arraybackslash\hspace{0pt}}m{#1}}

\makeatletter
\DeclareRobustCommand*\textsubscript[1]{%
  \@textsubscript{\selectfont#1}}
\def\@textsubscript#1{%
  {\m@th\ensuremath{_{\mbox{\fontsize\sf@size\z@#1}}}}}
\makeatother

\title{A topological criterion for filtering information in complex brain networks}

\author[1,2,*]{Fabrizio De Vico Fallani}
\author[3,4]{Vito Latora}
\author[2]{Mario Chavez}

\affil[1]{Inria, Aramis team, Paris, France}
\affil[2]{CNRS, UMR7225; Inserm, U1127; Sorbonne Universites, UPMC-P6 UMR S1127; ICM, Paris, France}
\affil[3]{Queen Mary University of London, School of Mathematical Sciences, London, UK}
\affil[4]{Universit\`a di Catania and INFN, Dipartimento di Fisica ed Astronomia, Catania, Italy}
\affil[*]{\scriptsize\textit{Corresponding author }}


\begin{abstract} 
In many biological systems, the network of interactions between the elements can only be inferred from experimental measurements. In neuroscience, non-invasive imaging tools are extensively used to derive either structural or functional brain networks in-vivo. As a result of the inference process, we obtain a matrix of values corresponding to an unrealistic fully connected and weighted network. To turn this into a useful sparse network, thresholding is typically adopted to cancel a percentage of the weakest connections. The structural properties of the resulting network depend on how much of the inferred connectivity is eventually retained. However, how to fix this threshold is still an open issue. We introduce a criterion, the efficiency cost optimization (ECO), to select a threshold based on the optimization of the trade-off between the efficiency of a network and its wiring cost. We prove analytically and we confirm through numerical simulations that the connection density maximizing this trade-off emphasizes the intrinsic properties of a given network, while preserving its sparsity. Moreover, this density threshold can be determined a-priori, since the number
of connections to filter only depends on the network size according to a power-law. We validate this result on several brain networks, from micro- to macro-scales, obtained with different imaging modalities. Finally, we test the potential of ECO in
discriminating brain states with respect to alternative filtering methods. ECO advances our ability to analyze and compare biological networks, inferred from experimental data, in a fast and principled way.
\end{abstract}

\begin{document}
%\flushbottom
\maketitle
\indent\small\textit{*  Institut du Cerveau et de la Moelle épiniere}\\
\indent\small\textit{  47, Boulevard de l'Hopital}\\
\indent\small\textit{  75013, Paris, France}\\
\indent\small\textit{  Email. fabrizio.devicofallani@gmail.com}\\
\noindent\normalsize
\thispagestyle{empty}
%
%\newpage

\section*{Introduction}

% 1) INTRODUCING IMAGING CONNECTOMICS
Imaging connectomics uses neuroimaging techniques to map connections and/or interactions between different brain sites. 
Combined with tools from graph theory, imaging connectomics has considerably advanced our understanding of the brain structure and function from a system perspective~\cite{bullmore_complex_2009}. 

Noninvasive neuroimaging is particularly attractive as it allows to map connectomes \textit{in-vivo} and quantify network organizational mechanisms underlying behavior, cognition, development as well as  disease~\cite{park_structural_2013, stam_modern_2014}. Magnetic resonance imaging (MRI) and electro/magnetoencephalography (E/MEG) are frequently used to derive macro-scale connectomes whose nodes, or units, correspond to spatially remote brain sites~\cite{craddock_imaging_2013}. High-field MRI and genetically encoded calcium indicators are promising tools to image connectomes respectively at the scale of neuronal ensembles (meso-scale) and single neurons (micro-scale) \cite{alivisatos_brain_2012}.
Brain connectivity methods are typically used to estimate the links between the nodes. While anatomical connectivity (AC) measures the probability to find axonal pathways between brain areas, typically from diffusion MRI, functional connectivity (FC) rather calculates the temporal dependence between their neural processes as recorded, for instance, by functional MRI, EEG or MEG ~\cite{bullmore_complex_2009}.

% 2) INTRODUCING THRESHOLDING PROBLEM 
At this stage, the resulting connectomes correspond to maximally dense networks whose weighted links code for the strength of the connections between different brain nodes. Common courses in brain network analysis use thresholding procedures to filter information in these raw connectomes by retaining and binarizing a certain percentage of the strongest links (\textbf{Supplementary Fig.~\ref{fig:SF1}}). 
Despite the consequent information loss, these procedures are often adopted to mitigate the incertainty carried by the weakest links and facilitate the interpretation of the network topology ~\cite{de_vico_fallani_graph_2014}. In addition, they enable to use all the graph theoretic tools, which are mainly conceived for sparse and binary networks ~\cite{boccaletti_complex_2006}.
At present, the choice of the specific value for such threshold remains arbitrary, so that scientists are obliged to explore brain network properties across a wide range of different candidates and eventually select one representative \textit{a-posteriori}~\cite{garrison_stability_2015}. These approaches are extremely time-consuming for large connectomes and make difficult the comparison between many individuals or samples~\cite{sporns_making_2013}. 

%AND OUR METHOD
We introduce a criterion to select \textit{a-priori} an optimal threshold which captures the essential topology of a connectome while preserving its sparsity. Based on the perfect trade-off between two desirable but incompatible features - namely high global and local integration between nodes, and low connection density - this method is inherently motivated by the principle of efficiency and economy observed in many complex systems \cite{latora_economic_2003}, including the brain \cite{bullmore_economy_2012}.

\section*{Results}
%nel testo dei results ho scelto di usare il past per escrivere cio che abbiamo fatto e il presente quando ci riferiamo a delle figure o interpretiamo i risultati. 

\subsection*{Filtering information as a network optimization problem} 
%piccola introduzione ma deve rimanere piccola
Global- and local-efficiency ~\cite{latora_efficient_2001} have revealed to be important graph quantities to characterize the structure of brain networks in terms of integration and segregation of information ~\cite{bassett_small-world_2006}.
Both anatomical and functional brain networks tend to exhibit relatively high values of global- and local-efficiency. At the same time they also tend to minimize, for economical reasons, the number of their links leading to sparse networks \cite{bullmore_economy_2012}. 

% descrizione dei risultati
Hence, we propose to determine a density threshold that filters out the weakest links and maximizes the ratio between the overall efficiency of a connectome and its wiring cost. 
We formally introduce a criterion to filter information in a given network by finding the optimal connection density $\rho$ that maximizes the quality function:
\begin{equation}
J=\dfrac{E_{g}+E_{l}}{\rho}\label{eq:1}
\end{equation} 
where $E_{g}$ and $E_{l}$ represent respectively the global- and local-efficiency of a network (Online Method 1). 

\begin{figure}[H]
%\begin{sidewaysfigure}[ht]
\renewcommand{\figurename}{Figure}
\centerline{\includegraphics[width=1.15\textwidth]{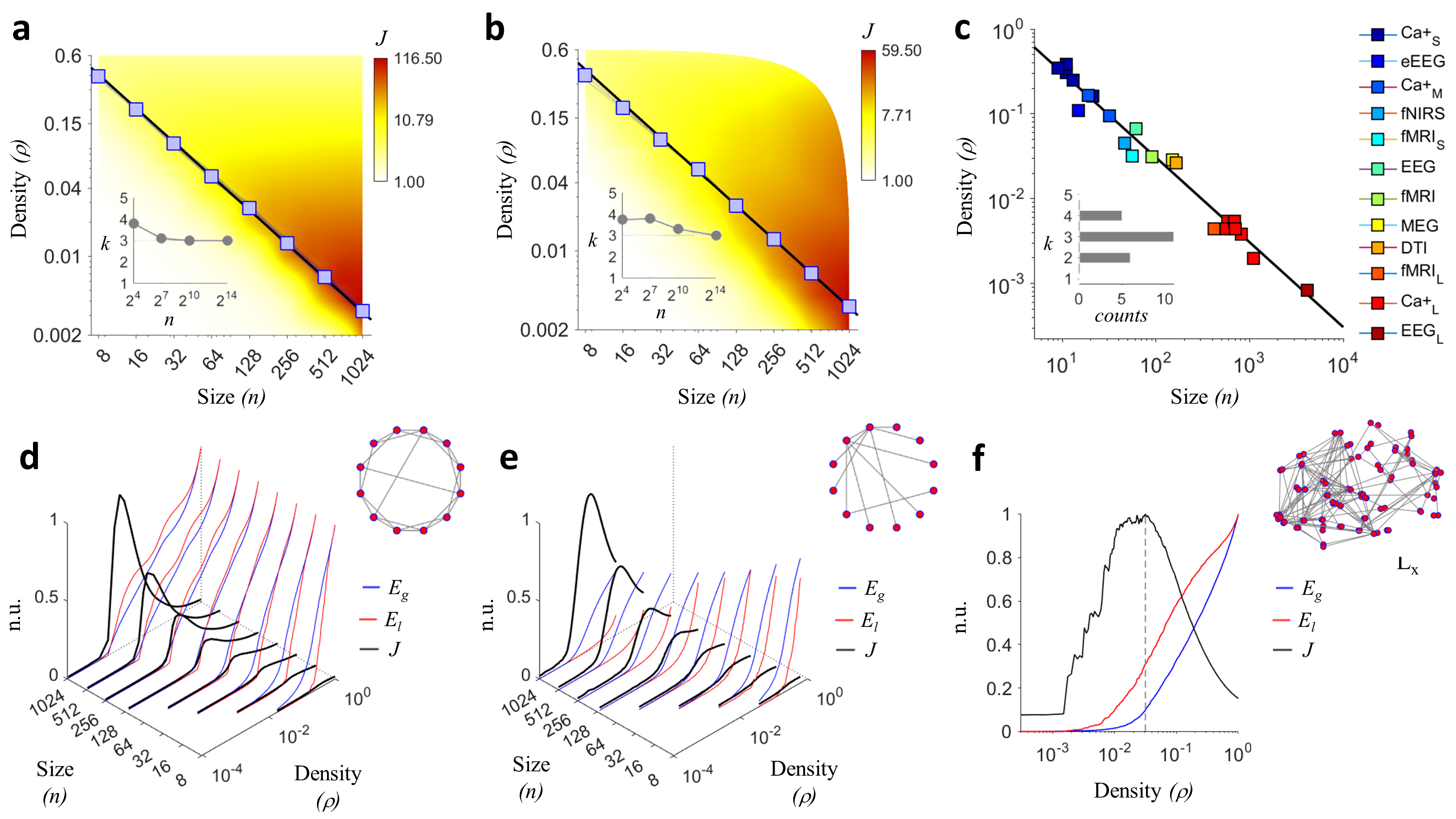}}
\caption{Optimal connection density in synthetic networks and brain networks. 
(\textbf{a}-\textbf{b}) Blue curves show the trends of the optimal density$\rho$ for one-hundred generated small-world $p_{ws}=0.1$ and scale-free $m_{ba}=9$ networks along different sizes $n$. Blue squares spot out the average optimal $\rho$ values. 
The black line shows the fit $\rho=c/(n-1)$ to the data, with $c=3.258$ for small-world networks and $c=3.215$ for scale-free networks (\textbf{Supplementary Table ~\ref{tab:ST1}}). 
The background color codes for the average value of the quality function $J$. Insets indicate that the optimal average node degree converges to $k=3$ for large network sizes ($n=16834$). 
(\textbf{c}) Optimal density values obtained from group-averaged $J$ profiles for different brain networks. Imaging connectomes come from previously published studies (\textbf{Table ~\ref{tab:1}}). The fit $\rho=c/(n-1)$ to the pooled data gives $c=3.06$ (adjusted $R^2=0.994$). The inset shows a sharp distribution for the optimal average node degree, with a mode $k=3$. 
(\textbf{d}-\textbf{e}) Average $J$ profile (black curves) for simulated small-world and scale-free networks as a function of the network size ($n$) and of the density ($\rho$). $J$ values are represented in normalized units (n.u.), having scaled them by the global maximum obtained for $n=1024$. Blue and red curves show respectively the profiles of global- ($E_g$) and local-efficiency ($E_l$).
(\textbf{f}) Group-averaged $J$ profile for fMRI connectomes (\textbf{Table ~\ref{tab:1}}). The grey dashed line indicates the actual density maximizing $J$, i.e., $\rho=0.035$, corresponding to an average node degree $k=3.115$. The graph illustrates the brain network of a representative healthy subject (lateral view, frontal lobe on the left $L_x$).
}\label{fig:1}
\end{figure}

For both regular lattices and random networks, we proved analytically that the optimal density deriving from the maximization of $J$ reads $\rho \simeq 3/(n-1)$, where $n$ is the network size, i.e., the number of nodes in the network (\textbf{Supplementary Text}). 
We confirmed this result (\textbf{Supplementary Fig. ~\ref{fig:SF2}a,b}) through extensive numerical simulations (Online Methods 2), showing that it held true also in more realistic network models, such as in small-world networks \cite{watts_collective_1998}  (\textbf{Fig. ~\ref{fig:1}a}) and in scale-free networks ~\cite{barabasi_emergence_1999} (\textbf{Fig. ~\ref{fig:1}b}). 
Notably, the optimal density emphasized the intrinsic structural properties of all the implemented synthetic networks in terms of global- and local-efficiency (\textbf{Fig. ~\ref{fig:1}d,e} and \textbf{Supplementary Fig. ~\ref{fig:SF2}d,e}).

\subsection*{Optimal density in connectomes derived from neuroimaging}
%io tendo ad usare connectome per riferirmi al dato non filtrato e brain network per il caso filtrato e una convenzione
We computed the quality function $J$ in both micro- and macro-scale brain networks and we evaluated how the optimal density scaled as a function of the network size. 
We considered connectomes used in previously published studies that were obtained with different imaging modalities, from calcium imaging to EEG, and constructed with different brain connectivity methods, from Pearson-correlation to Granger-causality (\textbf{Table ~\ref{tab:1}}). 

For each connectome we applied a standard density-based thresholding. We started with the empty network by removing all the links ($\rho=0$). Then, we reinserted and binarized one link at time, from the strongest to the weakest, until we obtained the maximally dense network ($\rho=1$).
At each step we computed $J$ and we recorded its profile as a function of $\rho$ (\textbf{Fig. ~\ref{fig:1}f}). 
If a study presented more samples (individuals) with connectomes of the same size, we considered the group-averaged $J$ profile to improve the quality of the estimation (\textbf{Supplementary Fig. ~\ref{fig:SF2}c,f}). 
The pooled density values, as returned by the maximization of $J$, followed the same scaling relationship that we reported for synthetic networks (\textbf{Fig.~\ref{fig:1}c}). This result confirms that also for brain networks we can assume that the optimal density threshold depends on the network size according to the same rule $\rho \simeq 3/(n-1)$.

In conclusion, we introduced a criterion, named efficiency cost optimization (ECO), to select a threshold leading to sparse, yet informative brain networks. Such a threshold does depend neither on how the connectome is constructed nor on its underlying structure, and can be therefore selected \textit{a-priori}.
\subsection*{ECO discriminated network properties of different brain states}
To illustrate the methodology, we considered connectomes from four different imaging modalities, namely EEG, MEG, fMRI, and DTI (\textbf{Table ~\ref{tab:1}}).
Because we do not know the true structure for these connectomes, we evaluated the ability of ECO to discriminate network properties of different brain states, i.e., healthy \textit{versus} diseased, at individual level.

We characterized brain networks by calculating graph quantities at different topological scales, i.e., large (global- and local-efficiency, $E_g$ and $E_l$), intermediate (community partition, $\mathpzc{P}$; and modularity, $Q$), and small (node degree, $k_i$; and betwenness, $b_i$) (Online Method 3).
To assess network differences between brain states, we measured distances between the values of the graph quantities obtained in the healthy group and those in the diseased group. We adopted the Mirkin index ($MI$) to measure distances between community partitions, and the divergent coefficient ($D$) for other graph quantities (Online Method 4).

\begin{figure}[H]
\renewcommand{\figurename}{Figure}
\centerline{\includegraphics[width=1.1\textwidth]{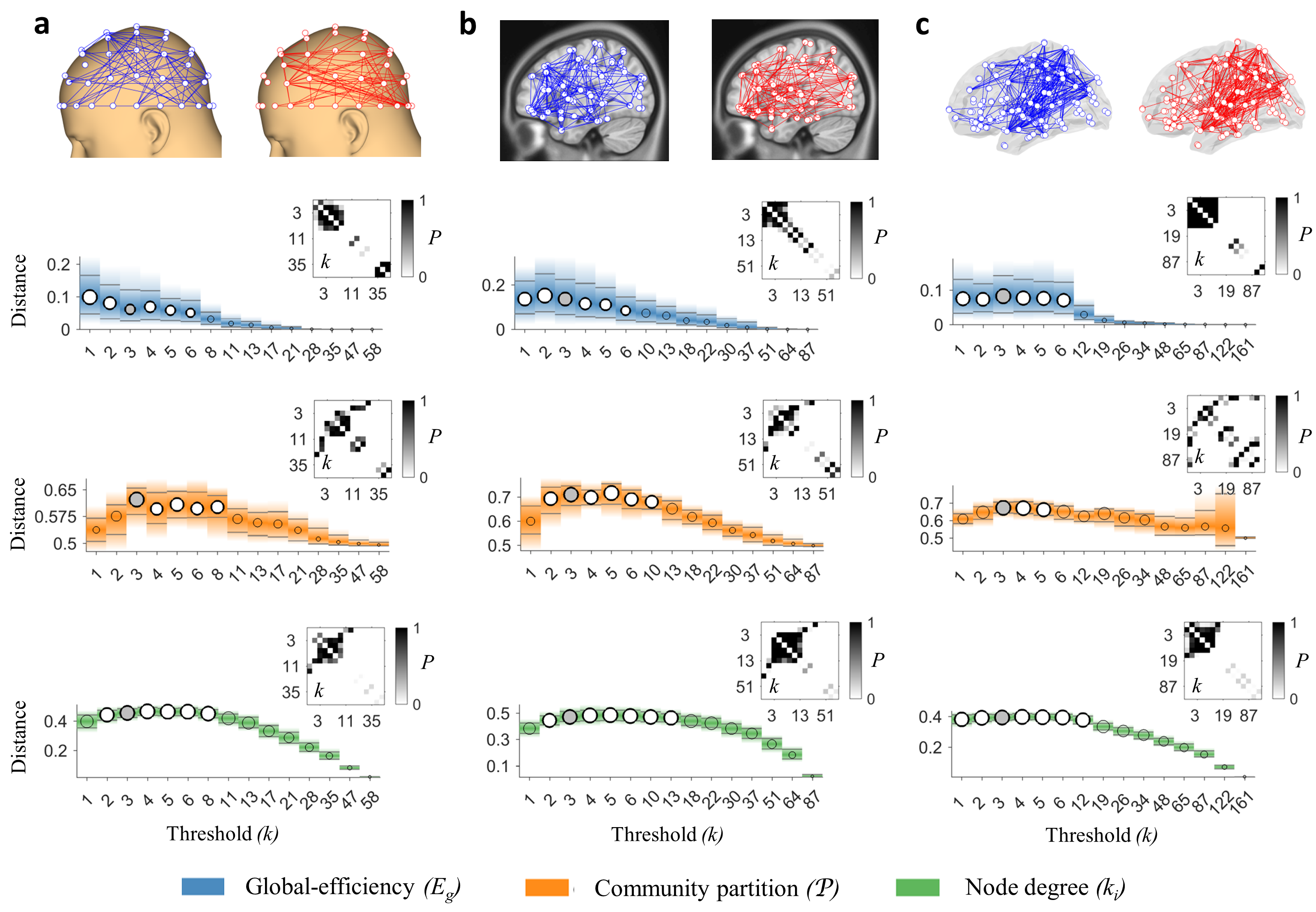}}
\caption{Statistical comparison of brain network distances across different thresholds.
(\textbf{a}-\textbf{c}) Top panels show group-averaged connectomes filtered with ECO for the healthy (blue links) and diseased (red links) group, in three representative imaging modalities, i.e., EEG, fMRI, and DTI (\textbf{Table ~\ref{tab:1}}). 
Lower panels show distances between brain network properties across different thresholds for global-efficiency $E_g$, community partition $\mathpzc{P}$, and node degree vector $\textrm{K}=[k_1, ..., k_n]$. Thresholds are given by the average node degree $k$, which corresponds to a connection density $\rho=k/(n-1)$. Circles correspond to medians; horizontal grey lines correspond to lower and upper quartiles; bar colors shade after quartiles. Overall, the distance significantly depends on the threshold value (Kruskalwallis test, $P<0.001$; \textbf{Supplementary Table ~\ref{tab:ST2}}). Grey circles represent distances for the optimal threshold $k=3$. White circles denote threshold values for which distances are not significantly different from $k=3$ (Tukey-Kramer post-hoc test, $P\geq0.001$). Transparent circles denote threshold values for which distances are significantly lower than $k=3$ (Tukey-Kramer post-hoc test, $P<0.001$). Insets show the $P$-values resulting from the Tukey-Kramer post-hoc comparison of distances between all the threshold values.
}\label{fig:2}
\end{figure}

\begin{figure}[H]
\renewcommand{\figurename}{Figure}
\centerline{\includegraphics[width=1.4\textwidth]{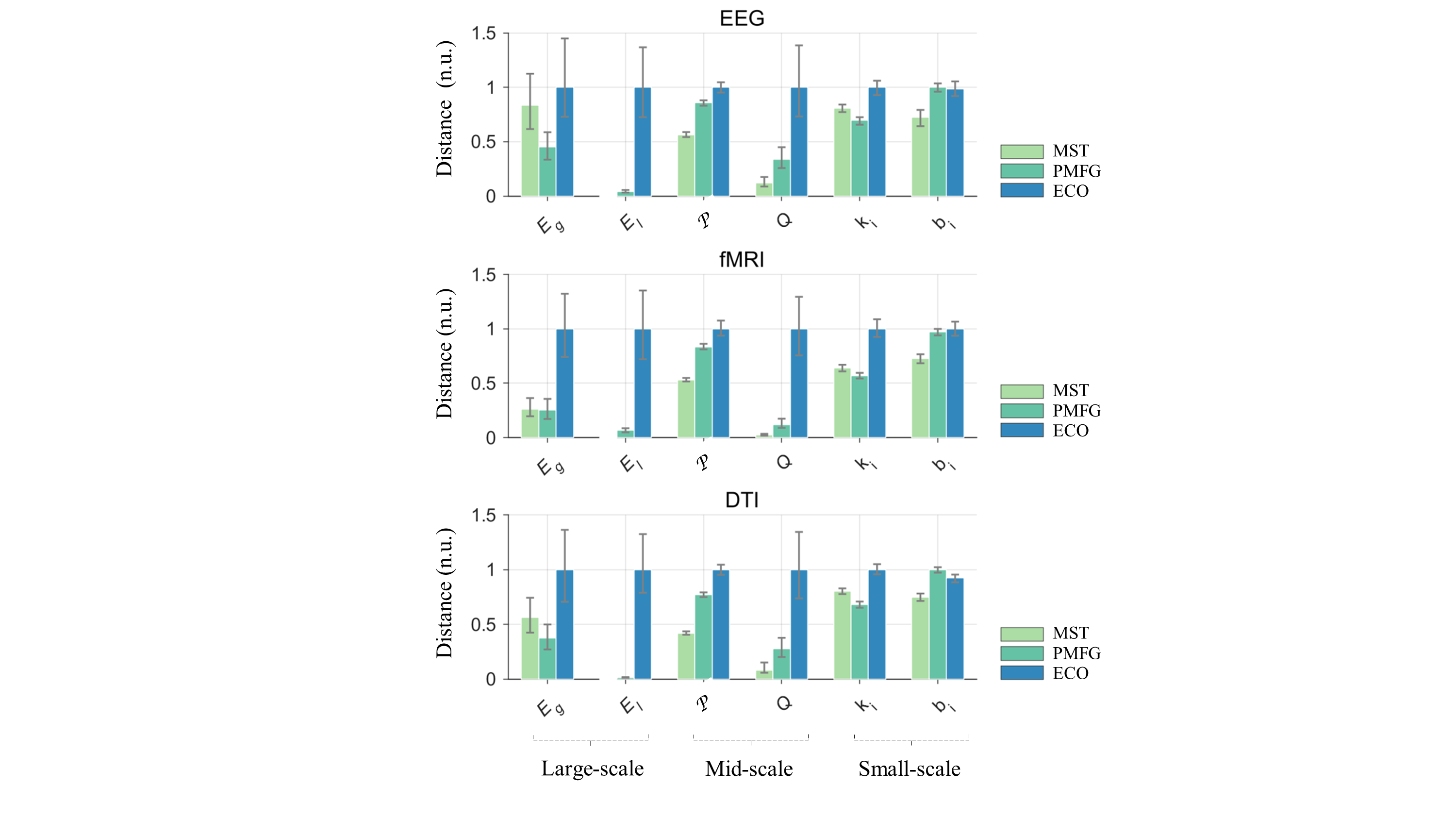}}
\caption{Statistical comparison of brain network distances across different filtering methods. Bar plots show the medians of distance between brain network properties of samples in the healthy and diseased group. Vertical bars denote lower and upper quartiles. Medians and quartiles are normalized for the sake of representation. Overall, the choice of the filtering method significantly affects distances between samples (Kruskalwallis test, $P<0.001$; \textbf{Supplementary Table ~\ref{tab:ST3}}). For all graph quantities, ECO gives significantly larger distances as compared to MST (Tukey-Kramer post hoc test, $P<0.01$) and, with minor extent, to PMFG (\textbf{Supplementary Table ~\ref{tab:ST4}}).
By construction, MST gives null distances for local-efficiency as there are no triangles in tree-like networks. 
}\label{fig:3}
\end{figure}

We explored a wide range of density thresholds and, as expected, the value of the threshold affected the ability to separate network properties of different brain states (\textbf{Fig.~\ref{fig:2}}, \textbf{Supplementary Fig.~\ref{fig:SF3} and ~\ref{fig:SF4}a}). Notably, the choice $\rho=3/(n-1)$ resulted among the best candidates in producing larger distances and improving discrimination. 
%modularity no
Furthermore, ECO overall outperformed alternative methods such as the minimum spanning tree (MST) and the planar maximally filtered graph (PMFG)~\cite{tumminello_tool_2005} (\textbf{Fig. ~\ref{fig:3}}, \textbf{Supplementary Fig. ~\ref{fig:SF4}b}, \textbf{Supplementary Table ~\ref{tab:ST2}}).

\section*{Discussion}

We introduced ECO to filter information in imaging connectomes whose links are statistical or probabilistic measures of brain connectivity. 
% 1) limitation of exisitng approaches -> advanteges and disadvantages of 3/N
Conventional graph approaches evaluate brain network properties across a large and arbitrary number of thresholds~\cite{fornito_graph_2013}. Eventually, they select a representative threshold \textit{a-posteriori} that maximizes the separation between different brain states ~\cite{de_vico_fallani_graph_2014}.
ECO provides a theoretically grounded criterion to select an optimal threshold \textit{a-priori}, drastically reducing the computational burden.
Other approaches, similar in purpose to ECO, impose unnatural constraints on the filtered connectome. The minimum spanning tree (MST), for instance, leads to brain networks with a null clustering coefficient ~\cite{tewarie_minimum_2015}. The planar maximally filtered graph (PMFG) tries to alleviate this bias by allowing closed loops, but still forces planarity~\cite{tumminello_tool_2005}.
Conversely, ECO does not impose any constraint and lets the intrinsic network structure of a connectome to emerge.

%2 mathematical issues; relmation to economic small world, alternatives
In addition, ECO is based on fundamental principles of complex systems~\cite{watts_collective_1998, barabasi_emergence_1999}.
Maximizing global- and local-efficiency with respect to connection density means emphasizing the integration and segregation properties of a connectome~\cite{tononi_measure_1994} while keeping a biologically plausible wiring cost.
This rationale dovetails with current evidence showing that advantageous topological properties, such as economic small-world architectures~\cite{bullmore_economy_2012}, tend to be maximized in brain networks, and that, in general, sparsity increases robustness of complex systems \cite{leclerc_survival_2008}.
Other combinations could have been considered when conceiving the quality function $J$. For example, in ~\cite{achard_resilient_2006} authors introduced the cost-efficiency $E_g - \rho$, which, however, did not include the clustering counterpart. That quality function, as well as other ones that we tested, did not exhibit an optimal density and was therefore not considered (\textbf{Supplementary Text}). 

%3 Final drawbacks
ECO makes use of density thresholds. Hence, filtered connectomes having same number of nodes, will have same number of links. On the one hand, this ensures that differences between brain network properties are not merely due to differences in the connection density~\cite{van_wijk_comparing_2010}.
On the other hand, ECO does not allow a direct evaluation of neural processes altering the number of links; yet it does inform on the consequent (re)organizational mechanisms. 
ECO is based on a graph topological criterion and cannot filter out possible false positives (i.e., spurious links) resulting from biased brain connectivity estimates~\cite{craddock_imaging_2013,de_vico_fallani_graph_2014}. This method assumes that the weighted links of the raw connectomes had been previously statistically validated, either maintained or canceled. 
%4) Conclusions, to be wide
We conceived ECO to filter imaging connectomes with applications ranging from cognitive to clinical and computational neuroscience. 
Given its generality and simplicity, we anticipate that ECO will facilitate the analysis of interconnected systems where the need of sparsity is plausible and links are weighted estimates of connectivity. 
This is, for example, the case of functional networks in system biology, where links are derived from transcriptional or phenotypic profiling, and genetic interactions~\cite{vidal_interactome_2011}.

\section*{Online Methods}

\subsection*{1. Topological properties of the quality function $J$} 
The proposed quality function can be seen as a particular case of a general family of functions of the form $f(E_g,E_l,\rho)$.
When combining different graph quantities, proper normalizing procedures need to be used~\cite{souza_optimization_2012}. 
By definition, each of the three quantities $E_{g}$, $E_{l}$ and $\rho$ is normalized in the range $[0,1]$, and $E_{g}$ and $E_{l}$ are non-decreasing 
functions of $\rho$. However, since the efficiency is based on shortest paths between nodes~\cite{latora_efficient_2001}, a concept which is not directly captured by the density, a scaling factor might be necessary to normalize changes among those quantities. 

We therefore considered a more general form of $J$ than that in Eq. \eqref{eq:1} by introducing two distinct dependencies on the connection density, i.e., $\beta\rho$ and $\rho^\beta$, where $\beta$ is a tunable control parameter.
For values around $\beta=1$, the parameter had no influence on the returned optimal density  (\textbf{Supplementary Text 3}).
Hence, for the sake of simplicity, we deliberately chose $\beta=1$ that corresponds to the original expression of the quality function in Eq. \eqref{eq:1}, i.e., $J=(E_{g}+E_{l})/\rho$.

When $\rho=0$ in Eq.\eqref{eq:1}, then both global- $E_{g}$ and local-efficiency $E_{l}$ are null leading to an indefinite form. As density slightly increases ($0<\rho<\epsilon$, with $\epsilon$ sufficiently small) it can be demonstrated that $J$ tends to $1$.
In fact, in this range, the probability to find at least three nodes connected together (a triangle) is extremely low. By definition, $E_{l}=0$ in absence of at least one triangle ~\cite{latora_efficient_2001} and therefore $J \simeq E_{g}/\rho$. By considering the definitions of $E_{g}$ and $\rho$, this quantity can be rewritten as
$E_{g}/\rho=1/m\sum_{i\neq j}^n 1/d_{i,j}$,
where $m$ is the number of existing links and $d_{i,j}$ is the distance between the nodes $i$ and $j$. In a generic network with $m$ links there are at least $m$ pairs of nodes directly connected (i.e., $d_{i,j}=1$). This means that the sum in the latter equation is bounded from below by $m$ in the case of isolated pairs of connected nodes ($m=n/2$) or in the trivial case of $m=1$. It follows that $J\rightarrow 1$ when there are relatively few links in a network.

When $\rho$ tends to $1$, it is trivial to see from equation \eqref{eq:1} that $J\rightarrow 2$, as both $E_g$ and $E_l$ tend to one. 
For intermediate density ranges ($\epsilon<\rho\ll1-\epsilon$) the analytical estimate of $J$ is not trivial since $E_g$ and $E_l$ depend on the network topology which is, in general, unknown.

\subsection*{2. Numerical simulations for small-world and scale-free networks}
Small-world networks were generated according to the Watts-Strogatz (WS) model ~\cite{watts_collective_1998} with a rewiring probability $p_{ws}=0.1$.
Scale-free networks were generated according to the Barabasi-Albert (BA) model ~\cite{barabasi_emergence_1999}.

In the first simulation, we considered undirected networks. 
We varied both the network size and the average node degree, i.e., $n=16, 128, 1024, 16384$ and $k={1, 2, 3, 4, 5}$. 
In the WS models, $k$ is even accounting for the number of both left and right neighbors of the nodes in the initial lattice. To obtain small-world networks with $k$ odd, we first generated lattices with $k$ even and then, for each odd node (e.g., $1$, $3$, ...), we removed the link with its left farthest neighbor. This procedure removes in total $n/2$ links leading to a new average node degree $k'=k-1$, while keeping a regular structure.
As for BA models, we set the number of links in the preferential attachment $m_{ba}=3$ and the initial seed was a fully connected network of $n_0=m_{ba}$ nodes. This setting generated scale-free networks with $k=6-12/n$, that is $k \geq 5$ regardless of the selected network size. We then removed at random the exceeding number of links until we reached the desired $k$ value.
This procedure had the advantage to preserve the original scale-free structure. 

In the second simulation, we considered directed networks to confirm and extend the results we obtained for undirected WS and BA networks. We selected eight representative network sizes, i.e., $n=8,16,32,64,128,256,512,1024$ covering the typical size of most current imaging connectomes, and we varied the connection density. Specifically, we performed a two-step procedure:
\begin{enumerate}
\item 
We fixed one-hundred $\rho$ values quadratically spaced within the entire available density interval.

\item 
After having identified the optimal $\rho^*$, we performed a refined research among one-hundred new values linearly spaced between the density values, in step 1, before and after $\rho^*$.
\end{enumerate}

For WS models, initial lattices had $k$ equal to the nearest even integer equal or higher than $\rho (n-1)$, with $\rho\in\left( 0, 1 \right)$.
For BA models, the number of attaching links was $m_{ba}=\log_2{n}$ to ensure an initial relatively high density; the seed was a fully connected network of $n_0=m_{ba}$ nodes. By construction
$\rho\in\left( 0 , \frac{2m_{ba}n+m_{0}}{n(n-1)}\right)$, where $m_{0}=n_0(n_0-1)/2$ is total number of links in the initial seed.
For both models, we then removed at random the exceeding links until we reached the desired density value.
For both simulation we generated one-hundred sample networks.

\subsection*{3. Graph analysis of brain networks}
Complex networks can be analyzed by a plethora of graph quantities characterizing different topological properties~\cite{costa_analyzing_2011}. Here, we considered a subset of representative ones which have been shown to be relevant for brain network analysis~\cite{rubinov_complex_2010}.
To characterize the entire brain network (i.e., large-scale topology), we used global- and local-efficiency, which respectively read:
\begin{align}
\label{eq:2}
\begin{split}
E_{g}&=\dfrac{2}{n(n-1)}\sum_{i\neq j}^n \dfrac{1}{d_{ij}}
\\
E_{l}&=\dfrac{1}{n}\sum_{i=1}^n E_g(i)
\end{split}
\end{align}
where $d_{ij}$ is the length of the shortest path between nodes $i$ and $j$, and $E_g(i)$ is the global-efficiency of the $i$th subgraph of the network ~\cite{latora_efficient_2001}.

To characterize modules, or clusters, of brain regions (i.e., mid-scale topology), we extracted the community partition $\mathpzc{P}$ of the brain network by means of the Newman's spectral algorithm maximizing the modularity:
\begin{equation}
Q=\frac{1}{2m}Tr(\mathbf{G^TMG})\label{eq:3}
\end{equation}
where \textbf{G} is the (non-square) matrix having elements $G_{ig}=1$ if node $i$ belongs to cluster $g$ and zero otherwise, and \textbf{M} is the so-called modularity matrix~\cite{newman_modularity_2006}.

To characterize individual brain areas (i.e., small-scale topology), we measured the centrality of the nodes in the brain network by means of the node degree and of the node betwenness, which respectively read:
\begin{align}
\label{eq:4}
\begin{split}
k_{i}&=\sum\limits_{j\neq i}^n A_{ij}
\\
b_{i}&=\sum\limits_{j\neq i \neq h} \dfrac{\sigma_{jh}(i)}{\sigma_{jh} }
\end{split}
\end{align}
where the element of the adjacency matrix $A_{ij}=1$ if there is a link between node $i$ and $j$, zero otherwise; and where $\sigma_{jh}$ is the total number of shortest paths between nodes $j$ and $h$, while $\sigma_{jh}(i)$ is the number of those paths that pass through $i$.

\subsection*{4. Distances between samples and statistical analysis}
To assess brain network differences between individuals (or samples) in the two groups, we measured the distance between the respective values obtained for each graph quantity.
We used the Mirkin index to compute distances between two network partitions $\mathpzc{P}_u$ and $\mathpzc{P}_v$: 
\begin{equation}
MI(\mathpzc{P}_u, \mathpzc{P}_v)=2(n_{uv} + n_{vu}) \label{eq:5}
\end{equation}
where $n_{uv}$ is the number of pairs of nodes that are in different clusters under $\mathpzc{P}_u$  but not in the same one under $\mathpzc{P}_v$; and $n_{vu}$ is the number of pairs that are in the same cluster under $\mathpzc{P}_u$ but in different ones under $\mathpzc{P}_v$ ~\cite{meila_comparing_2007}. 
For all other graph quantities, we used the divergent coefficient~\cite{cox_multidimensional_2008}:
\begin{equation}
D(\textrm{X}_u, \textrm{X}_v)=\sqrt{\left(\frac{1}{M}\sum\limits_{m=1}^M \frac{x_{u,m}-x_{v,m}}{x_{u,m}+x_{v,m}}\right)^2}\label{eq:6}
\end{equation}
where $\textrm{X}_u=[x_{u,1}, x_{u,2}, ..., x_{u,M}]$ and $\textrm{X}_v=[x_{v,1}, x_{v,2}, ..., x_{v,M}]$, contain the value(s) of the graph quantity for the $u$th and $v$th sample. Notably, $M=1$ for global-, local-efficiency and modularity (i.e., $E_g$, $E_l$, $Q$). $M=n$ for the node degree vector $\textrm{K}=[k_1, k_2, ..., k_n]$ and the node betweenness vector $\textrm{B}=[b_1, b_2, ..., b_n]$.

We used Kruskal–Wallis one-way analysis of variance to evaluate the overall effect of different thresholds, or filtering methods (i.e., MST, PMFG) on distances between individuals. 
A Tukey-Kramer multiple comparison test was then used to determine specific differences between pairs of thresholds or methods~\cite{mcdonald_handbook_2009}. 

\section*{Acknowledgements}

We thank all the colleagues who shared their databases (see \textbf{Table \ref{tab:1}}) allowing to validate the proposed method. We are grateful to Sophie Achard, Vincenzo Nicosia, Jonas Richiardi, and Miguel Valencia for their useful comments and suggestions. V.L. and M.C. acknowledge support by the European Commission Project LASAGNE Grant 318132; V.L. acknowledges support from EPSRC project GALE Grant EP/K020633/1; F.D. and M.C. acknowledge support by French program "Investissements d'avenir" ANR-10-IAIHU-06; F.D. acknowledges support by the project NETBCI Grant ANR-15-NEUC-0006-02.

\section*{Author contributions statement}

F.D. conceived the study and performed data analysis; V.L. and M.C. contributed to the analytical proofs and simulations. All authors wrote and reviewed the manuscript.

\section*{Competing financial interests}
The authors declare no competing financial interest.

\section*{Materials and correspondence}
Specific inquiries on experimental data can be addressed to the corresponding authors listed in \textbf{Table \ref{tab:1}}.

\bibliography{Mybib_20032016}

%\noindent LaTeX formats citations and references automatically using the bibliography records in your .bib file, which you can edit via the project menu. 

\newpage
%\begin{landscape}
%%% FIGURES AND TABLES
\section*{Figures}

%\end{landscape}

\newpage

%\end{landscape}

\newpage

\newpage
\section*{Tables}

\begin{table}[H]
%begin{sidewaystable}[ht]
%\scriptsize
\footnotesize
%\centering
%\resizebox{\textwidth}{!}{%
\hspace{-0.75cm}
\begin{tabular}{L{1.4cm} L{1.4cm} l L{1.25cm} L{2cm} l L{2.25cm} L{2cm} L{3cm}}
%& & Experimental data & & & & Connectome specifics & &                \\
\textbf{}  \textbf{Imaging modality}    & \textbf{Group(s)}      & \textbf{Species}  & \textbf{Samples x Group} & \textbf{Condition} & \textbf{Nodes} & \textbf{Connectivity method} & \textbf{Domain}            & \textbf{Links} 
\\

\cite{de_vico_fallani_hierarchy_2014} Ca\textsuperscript{+}\textsubscript{S} & Healthy & Zebrafish & 5 & Spontaneuous & {[}9,21{]} & Granger causality & Time & Directed       \\
\cite{plomp_physiological_2014} eEEG      & Healthy                        & Rodent            & 1                        & Evoked potential   & 15                 & Partial directed coherence   & Time/Freq. (8 ms/14-29 Hz) & Directed       \\
\cite{teller_emergence_2014} Ca\textsuperscript{+}\textsubscript{M}      & -                     & Culture           & 2                        & Spontaneous       & {[}19,32{]}      & Time delay                   & Time                       & Directed       \\
\cite{niu_test-retest_2013} fNIRS     & Healthy             & Human             & 2                        & Resting state      & 46                 & Pearson's correlation       & Time                       & Undirected     \\
\cite{oreilly_causal_2013} fMRI\textsubscript{S}     & Healthy             & Primate           & 3                        & Resting state      & 56                 & Pearson's correlation       & Time                       & Undirected     \\
\cite{de_vico_fallani_multiscale_2013} EEG      & Healthy, Stroke        & Human             & 20                       & Motor imagery      & 61                 & Imaginary coherence          & Frequency (14-29 Hz)       & Undirected \\
\cite{achard_hubs_2012} fMRI     & Healthy, Coma      & Human             & 17                       & Resting state      & 90                 & Wavelet correlation          & Time                       & Undirected    \\
\cite{chavez_functional_2010} MEG       & Healthy, Epilepsy      & Human             & 5                        & Resting state      & 149                & Spectral coherence           & Frequency (5-14 Hz)        & Undirected   \\
\cite{besson_structural_2014} DTI       & Healthy, Epilepsy  & Human             & 19                       & -                  & 164                & Fractional anisotropy        & -                          & Undirected     \\
\cite{achard_hubs_2012} fMRI\textsubscript{L}     & Healthy, Coma      & Human             & 17                       & Resting state      & 417                & Wavelet correlation          & Time                       & Undirected    \\
\cite{teller_emergence_2014} Ca\textsuperscript{+}\textsubscript{L}     & -                      & Culture           & 6                        & Spontaneous       & {[}562,1107{]}   & Time delay                   & Time                       & Directed      \\
\cite{de_vico_fallani_community_2012} EEG\textsubscript{L}      & Healthy                     & Human             & 5                        & Motor execution    & 4094               & Imaginary Coherence          & Frequency (13-30 Hz)       & Undirected   
\end{tabular}

%}
\caption{Experimental details and network characteristics of imaging connectomes.
}
\label{tab:1}

\end{table}

%\end{landscape}
%}

\end{document}